# Self-Powered, Ultra-thin, Flexible, and Scalable Ultraviolet Detector Utilizing Diamond-MoS$_2$ Heterojunction

*Yicheng Wang‡, Jixiang Jing‡, Yumeng Luo‡, Xiaomin Wang, Kuan Liang, Changsheng Chen, Dong-Keun Ki, Ye Zhu, Zhongqiang Wang, Qi Wang, Kwai Hei Li\*, Zhiqin Chu\**

‡ These authors have contributed equally to this work.
\* Corresponding authors

Yicheng Wang, Jixiang Jing, Zhiqin Chu,

Department of Electrical and Electronic Engineering, The University of Hong Kong, Pokfulam, Hong Kong.

E-mail: zqchu@eee.hku.hk

Yumeng Luo, Kwai Hei Li

School of Microelectronics, Southern University of Science and Technology, Shenzhen 518055, China.

E-mail: khli@sustech.edu.cn

Zhongqiang Wang, Qi Wang

Dongguan Institute of Opto-Electronics, Peking University, Dongguan 523808, China.

Xiaomin Wang, Dong-Keun Ki

Department of Physics, The University of Hong Kong, Pokfulam, Hong Kong.

Kuan Liang, Changsheng Chen, Ye Zhu

Department of Applied Physics, The Hong Kong Polytechnic University, Hong Kong.

**Abstract**: The escalating demand for ultraviolet (UV) sensing in space exploration, environmental monitoring, and agricultural productivity necessitates detectors that are both environmentally and mechanically resilient. Diamond, featuring its high bandgap and UV absorption, exceptional mechanical/chemical robustness, and excellent thermal stability, emerges as a highly promising material for next-generation UV detection in various scenarios. However, conventional diamond-based UV detectors are constrained by rigid bulk architectures and reliance on external power supplies, hindering their integration with curved and flexible platforms and complicating device scalability due to auxiliary power requirements. To tackle these challenges, herein, we firstly demonstrated a large-scale, self-powered, and flexible diamond UV detector by heterogeneously integrating a MoS$_2$ monolayer with an ultrathin, freestanding diamond membrane. The fabricated device operates at zero external bias, and simultaneously exhibits a high responsivity of 94 mA·W$^{-1}$ at 220 nm, and detectivity of 5.88 × 10$^9$ Jones. Notably, mechanical bending enables strain-induced bandgap modulation of the diamond membrane, allowing dynamically tunable photoresponse—a capability absent in rigid diamond counterparts. To validate its practicality and scalability, a proof-of-concept UV imager with 3×3 pixels was demonstrated. This newly developed configuration will undoubtedly open up new routes toward scalable, integrable, flexible, and cost-effective UV sensing solutions for emerging technologies

Ultraviolet (UV) radiation has a variety of impacts on human health,[1,2] ecosystems,[3] and agricultural productivity. [4,5] Precise monitoring of UV radiation light is of great significance for medical and biological research, space exploration, and chemical analysis. Thus, developing high-performance, low-cost, stable and durable UV photodetectors is crucially important. [6-10] Although reported UV detectors exhibit outstanding sensing performance, the lack of resistance to chemical corrosion and radiation makes them incompatible in harsh environments such as acid/alkali conditions or outer space.

Diamond emerges as an ideal candidate for robust UV detection due to its exceptional chemical stability, ultra-high bandgap, high carrier mobility, outstanding thermal conductivity, and low dielectric constant. However, inefficient separation of photogenerated electron-hole pairs makes diamonds heavily rely on an external electric field to achieve practical responsivity, which necessitates an external power supply and complicates device integration.[11-14] In contrast, self-powered photodetectors, based on the built-in electric field of p-n junctions, [15-17] heterojunctions, [18-20] Schottky junctions, [21-23] and other hybrid junctions, [24-26] present viable alternatives. [27-29] Moreover, most existing diamond detectors are fabricated from rigid, expensive bulk materials with limited scalability, [30-33] and their inflexibility impedes integration with curved/flexible platforms. Additional challenges, including limited responsivity, inadequate signal output, and complex fabrication, hinder broader adoption in next-generation UV sensing applications. Thus, Flexible, self-powered diamond UV detectors represent an urgent technological need to address aforementioned problems.

Recently, the scalable and flexible diamond membranes have been realized by our proposed edge-exposed exfoliation method. [34] Based on this platform, we herein demonstrate a flexible, self-powered diamond – molybdenum sulfide ($MoS_2$) heterojunction UV detector. $MoS_2$ was chosen for its inherent flexibility compatible with diamond membranes, and the ability to form clean van der Waals interfaces without high-temperature processing. [35-38] The fabricated device achieves high-performance operation under zero bias, exhibiting a fast response time below 40 ms, a responsivity of 94 mA·$W^{-1}$, and a detectivity of $5.88 \times 10^9$ Jones under 220 nm UV illumination. Moreover, Mechanical bending of the diamond membrane enables bandgap engineering for tunable photoresponse, enhancing responsivity by up to 117.9%. A proof-of-concept 3×3 pixel array further validates the scalability and potential for solar-blind imaging applications. These findings highlight a new approach to developing flexible, durable, and high-performance diamond-based UV detectors for emerging applications, such as astronomical sensing, UV imaging, and wearable optoelectronic devices.

**Results and discussion**

**Characterizations of Diamond – MoS2 heterojunction**

Conventional polycrystalline diamond films exhibit inherent surface irregularities arising from grain anisotropy and random crystallographic orientations, with roughness values scaling directly with film thickness. These topological limitations fundamentally impede their compatibility with 2D materials. In contrast, diamond membrane used in this work provides an ultra-smooth surface morphology with sub-nanometer roughness (Ra $\approx$ 1 nm), enabling robust and seamless integration with 2D materials (Figure S1). Figure 1(a) presents a schematic illustration of flexible diamond – MoS$_2$ UV detector. The device is constructed by transferring a monolayer MoS$_2$ onto a 5-μm-thick diamond membrane, which is supported by a polydimethylsiloxane (PDMS) substrate. Two gold electrodes with a thickness of 50 nm, patterned via photolithography and electron beam deposition, establish contact with the MoS$_2$ layer and diamond regions. This design leverages the inherent flexibility of both the diamond membrane and 2D materials, enabling conformal integration with various substrates. The mechanical flexibility of the fabricated detector is demonstrated in Figure 1(b), where a 1.5 × 2.5 cm² diamond membrane is conformally attached to a bent PDMS substrate. The scalable platform supports extension to wafer-scale dimensions.

Interface quality was examined in Figure 1(c), which shows a magnified microscope image at the diamond – MoS$_2$ interface. The inset plots height profile characterized by atomic force microscope (AFM), indicating that MoS$_2$ monolayer has a thickness of approximately 0.65 nm. Figure 1(d) shows Raman shift of bare diamond and MoS$_2$ on its as-grown substrate, where a prominent peak at 1332 cm$^{-1}$ indicates typical diamond Raman status of the membrane, and two typical peaks at 384 cm$^{-1}$ and 403 cm$^{-1}$ correspond to the vibration modes $E^2g^1$ and $A_1g$ of monolayer MoS$_2$, respectively. [39,40] The Raman shift exhibits similar characteristics after transferring MoS$_2$ onto diamond, as shown in Figure 1(e). Figure 1(f) and (g) exhibit cross-sectional images of diamond-MoS$_2$ interface taken with transmission electron microscope and energy dispersive X-ray microscopy (EDS), respectively. MoS$_2$ monolayer is highlighted with the ball-and-stick model in Figure 1(f). The EDS mapping in Figure 1(g) further confirms the successful integration of MoS$_2$, where carbon and Platinum layer were deposited on top of MoS$_2$ as protective layer for the subsequent focused ion beam treatment.

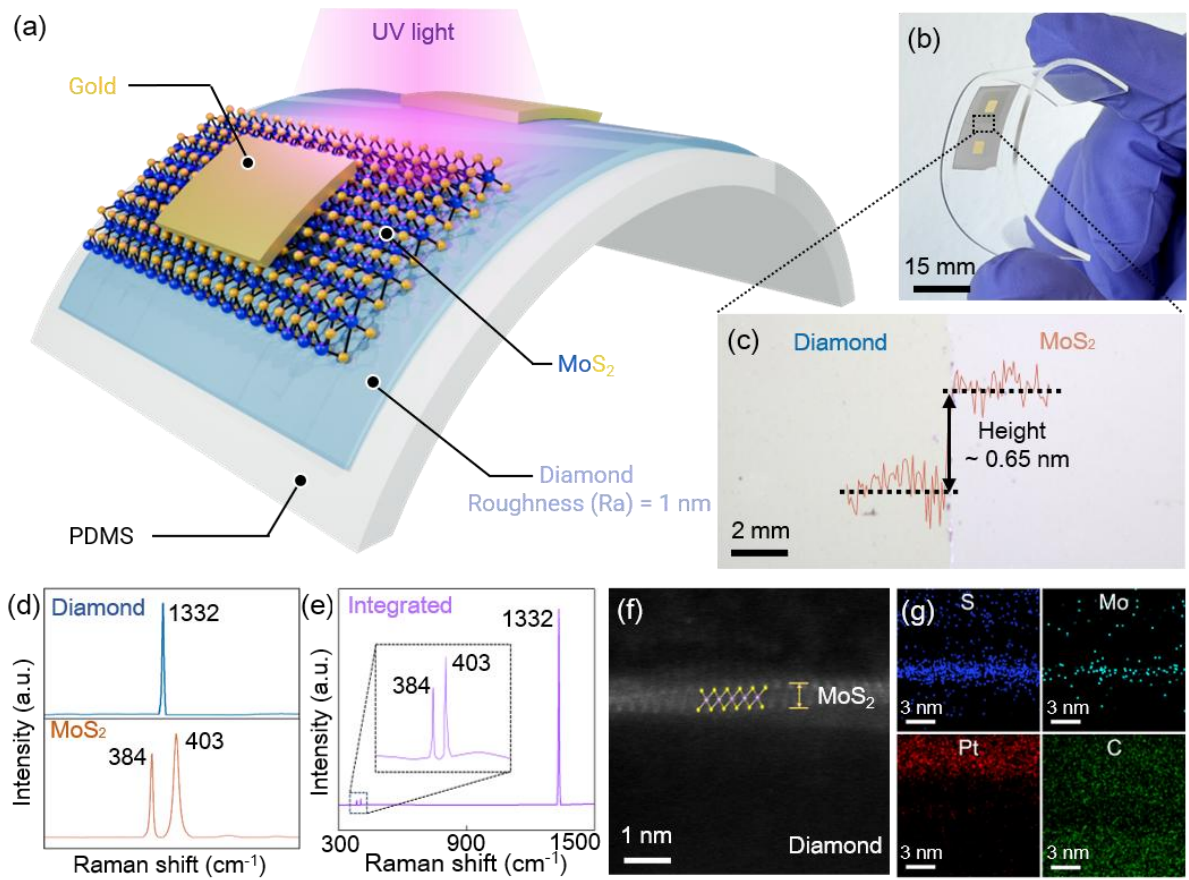

**Figure 1.** (a) Schematic illustration of the flexible diamond – MoS$_2$ UV detector. (b) Photograph of the flexible diamond UV detector. (c) Microscope image of the diamond / MoS$_2$ interface, inset: height difference at the interface measured by AFM. (d) Raman spectra of the diamond and MoS$_2$. (e) Raman spectrum of the diamond integrated with MoS$_2$. Cross-sectional images of the

diamond-MoS$_2$ interface taken with (f) transmission electron microscope and (g) energy dispersive X-ray microscopy.

**UV detection performance of the diamond – MoS$_2$ heterojunction**

Figure 2(a) depicts the experimental setup for characterizing the proposed UV detector, where the diamond – MoS$_2$ integration is mounted on a flat PDMS substrate, and an ammeter is connected to both electrodes for photoresponse measurements. As shown in Figure 2(b), the diamond – MoS$_2$ heterojunction shows a remarkable enhancement in responsivity (~10$^4$ times) over the bare diamond membrane, which is attributed to the fast separation of photogenerated electron-hole pairs by the built-in electric field of the heterojunctions. Figure 2(c) presents I-V curves of the proposed UV detector under 220-nm UV illumination at a varied intensity ranging from 0 to 500 µW·cm$^{-2}$. Under 200 µW·cm$^{-2}$ illumination and zero bias, the detector exhibits a photocurrent of 4.7 µA. To evaluate the satisfying responsivity ($R$) of the proposed UV detector, I-V plots were obtained under 200 µW·cm$^{-2}$ intensity at different wavelengths ranging from 220 nm to 300 nm, as shown in Figure 2(d). The responsivity is calculated using the following equation: [41]

$$R = \frac{I_{ph}}{P \cdot S} = \frac{I_{light} - I_{dark}}{P \cdot S} \quad (1)$$

where $I_{light}$, $I_{dark}$, $P$, $S$ represent the current under illumination, current in the dark condition, light power intensity, and effective area of the photodetector, respectively. The responsivity of the proposed UV detector gradually increases from 7.54 mA·W at 300 nm to 94.07 mA·W at 220 nm. The detectivity ($D$), which is used to describe the minimum detectable signal, can be calculated as follows: [42]

$$D = \frac{\sqrt{S} \cdot R}{\sqrt{2eI_{dark}}} \quad (2)$$

where $e$ is the electron charge. The proposed detector achieves a good detectivity of $5.88\times10^9$ Jones. Figures 2(e) and (f) show the time-dependent photoresponse of the device at zero bias under intermittent 220nm UV illuminations of different durations and intensities. The photoresponse also remains stable and repeatable throughout the test cycles. Additionally, Figure 2(g) presents the fast response of the device, with a rise time of approximately 40 ms. Another important parameter evaluating the minimum photon power required to overcome noise power is noise equivalent power (NEP), which could be calculated as follows:

$$NEP = \frac{\sqrt{i_n^2}}{R} \qquad (3)$$

where $i_n$ is the time averaged noise power, and $R$ represents responsivity, the calculated NEP for the proposed detector is $1.702 \times 10^{-10}$ W for 220 nm incident light, demonstrating practical reliability.

Compared with conventional diamond UV detectors, the proposed diamond membrane UV detector possesses not only improved responsivity, fast response, and self-powered operation, but also mechanical flexibility that is unattainable in bulk diamond. Key performance metrics are summarized in Table. 1.

**Table 1.** Comparison of diamond-based photodetectors

| Responsivity (mA·W$^{-1}$) | Response time | Self-powered | Flexibility | Scalability | Reference |
|---|---|---|---|---|---|
| 94@220nm, 0V bias | 40 ms | √ | √ | 2 cm[a] | This work |
| 0.2@270nm, 0V bias | 1 s | √ | × | - | [6] |
| 3.76×10$^{-4}$@225nm, 3V bias | - | × | √ | 2 cm | [29] |
| 45@228nm, 5V bias | 20 μs | × | × | 2 inches | [30] |
| 2.29@210nm, 5V bias | 800 ms | × | × | 10 μm | [32] |
| 5@193nm, 0V bias | 3 s | √ | × | 5 mm | [33] |
| 0.0017@210nm, 10V bias | - | × | × | 400 μm | [43] |
| 200@220nm, 5V bias | - | × | × | - | [44] |
| 3@225nm, 8V bias | 60 ms | × | × | - | [45] |
| 3.71@215nm, 5V bias | 590 ms | × | × | 3 mm | [46] |
| 2.94@215nm, 20V bias | - | × | × | 3 mm | [47] |

a). (In principle, the scalability could be extended to 4 inches.)

The self-powered characteristics originate from the energy band of a diamond – MoS$_2$ heterojunction. The energy band diagrams before and after contact are plotted in Figure 2(h) and (i), respectively. The bandgap of the polycrystal diamond membrane and monolayer MoS$_2$ are 4.4 eV and 2.3 eV, respectively, while their electron affinity are 1.7 eV and 4 eV. Therefore, the conduction band offset ($\Delta Ec$) and valance band offset ($\Delta Ev$) at the interface can be calculated to be 2.3 eV and 0.2 eV, respectively. Upon contact, electrons flow from the diamond to MoS$_2$ until the equilibrium of Fermi levels across the junction is reached. This charge transfer results in the formation of a depletion region and induces downward band bending in the diamond, further establishing a built-in electric field near the interface. Under UV illumination, photogenerated electron-hole pairs are rapidly separated, and their recombinations are suppressed. The electrons transfer towards MoS$_2$ and holes transfer toward the diamond, resulting in the high responsivity and self-powered performance.

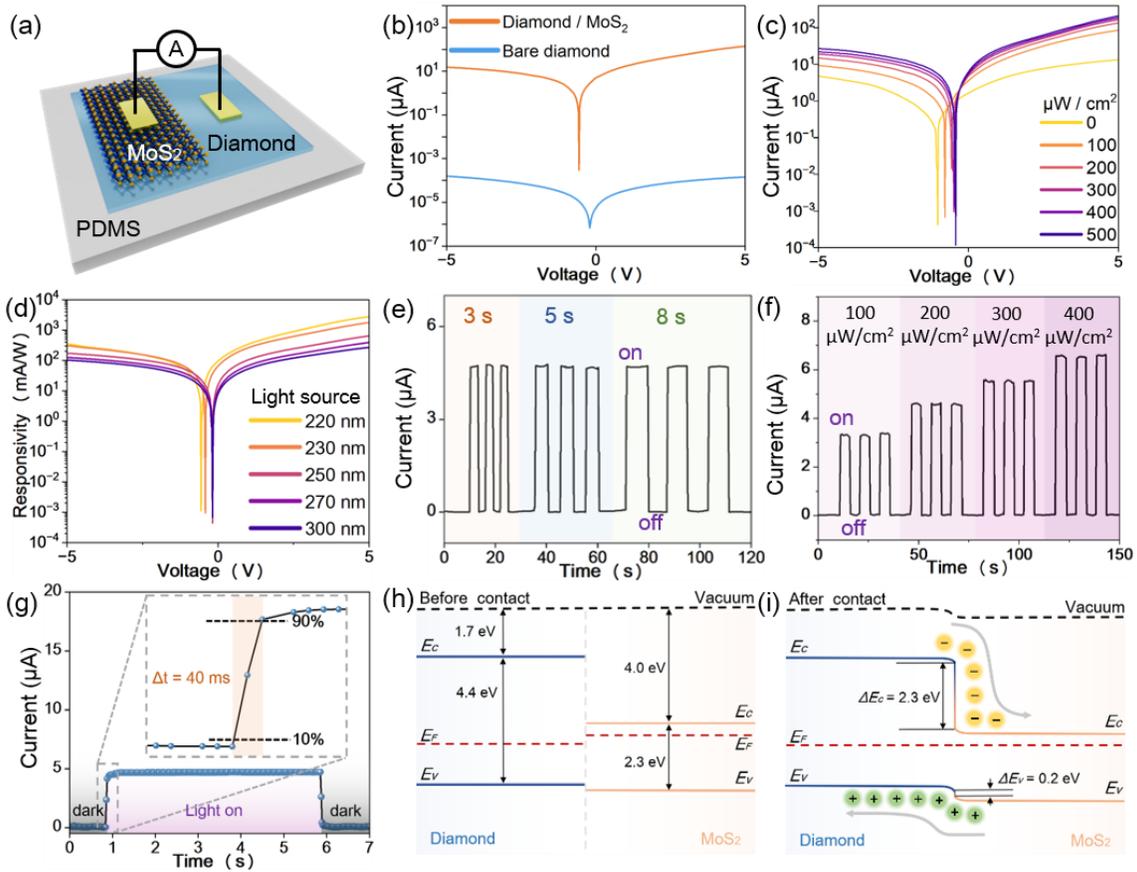

**Figure 2.** (a) Schematic diagram indicating the measurement of the diamond UV detector under unstrained conditions. (b) I-V curves of the bare diamond and diamond-MoS$_2$ heterogeneous device. (c) I-V plot of diamond-MoS$_2$ UV detector under 220-nm UV illumination with varying intensities. (d) Responsivity of the diamond UV detector under different wavelengths of illumination. Photoresponse of the diamond UV detector during on/off loops of UV illumination with different (e) durations and (f) illumination intensities. (g) Response time of the diamond UV detector. Energy band illustration of diamond-MoS$_2$ heterojunction (h) before contact and (i) after contact.

**Strain enhanced responsivity of flexible diamond UV detector**

The inherent flexibility of both the diamond membrane and 2D materials enables conformal integration of the UV detector with various curved surfaces. The influence of mechanical strain on device performance is evaluated by using a PDMS substrate subjected to controlled bending. Figure 3(a) schematically depicts the experimental setup, where the bending angle $\theta$ quantifies deformation applied to the diamond membrane. Figure 3(b) presents I-V curves of the flexible UV detector under 220 nm, 200 µW·cm$^{-2}$ illumination intensity at varying values of $\theta$. While the substrate is bent to $\theta = 5°$ and $\theta = 10°$, the photoresponse increases by 7.8% and 17.9%, respectively. In addition, Figure 3(c) shows the time-dependent photoresponse of the detector under intermittent 220 nm UV illuminations at zero bias with different values of $\theta$. These bending experiments replicate the enhancement trends observed in Figure 3(b), confirming the strain-induced enhancement effect. Notably, despite anticipated photocurrent reduction from decreased effective illumination area during bending, the observed response increase indicates the dominant influence of strain-engineered carrier dynamics over geometric effects.

In order to further understand the mechanism of the adjustable photoresponse, light absorption spectra of both bare diamond and diamond – MoS$_2$ heterojunction were characterized at bending angle $\theta$ from 0° to 15°, as shown in Figure 3(d) and (e), respectively. Both configurations exhibit progressive absorption enhancement with increasing $\theta$, attributed to the bandgap narrowing in the diamond membrane under tensile strain. This relationship was quantified by converting the absorption spectra into a Tauc plot; the bandgaps ($E_g$) are acquired from the following Tauc equation [48]:

$$(Ah\nu)^n = c(hc - E_g) \tag{4}$$

where $h$ is Planck constant, $v$ denotes the frequency of the incident electromagnetic wave, $A$ denotes the absorption coefficient, and $c$ is the proportionality coefficient. For semiconductors with direct and indirect bandgap, the value of $n$ is 0.5 and 2, respectively. As shown in Figure 3(f), the calculated bandgaps of the bare diamond and diamond- $MoS_2$ heterostructure decrease by 0.15eV and 0.155eV, respectively, as $\theta$ increases from 0° to 15°. Although mechanical strain can increase the risk of fracture, particularly due to the polycrystalline nature and defect density in the diamond membrane, previous studies have shown that such membranes can endure tensile strains up to 4.08% and withstand over 10,000 deformation cycles without failure, [34] with all experiments herein conducted within these operational limits.

The trend of bandgap narrowing of the diamond membrane coincides with previous works, [49,50] where a bandgap down to 0.62 eV was observed experimentally under tensile strain at 9.6%. Under tensile stress, the lattice of diamond elongates, resulting in an increase in the C-C bond length and a weakening of the electronic interactions between carbon atoms. This bond lengthening reduces electron cloud density, lowering the conduction band minimum and narrowing the bandgap. Theoretical prediction suggests the bandgap of the diamond would decrease to zero with tensile strain exceeding 12%, [50] indicating the possibility of indirect-to-direct bandgap transition of diamond. The combination of performance tunability and mechanical robustness makes this platform particularly attractive for flexible UV sensing applications.

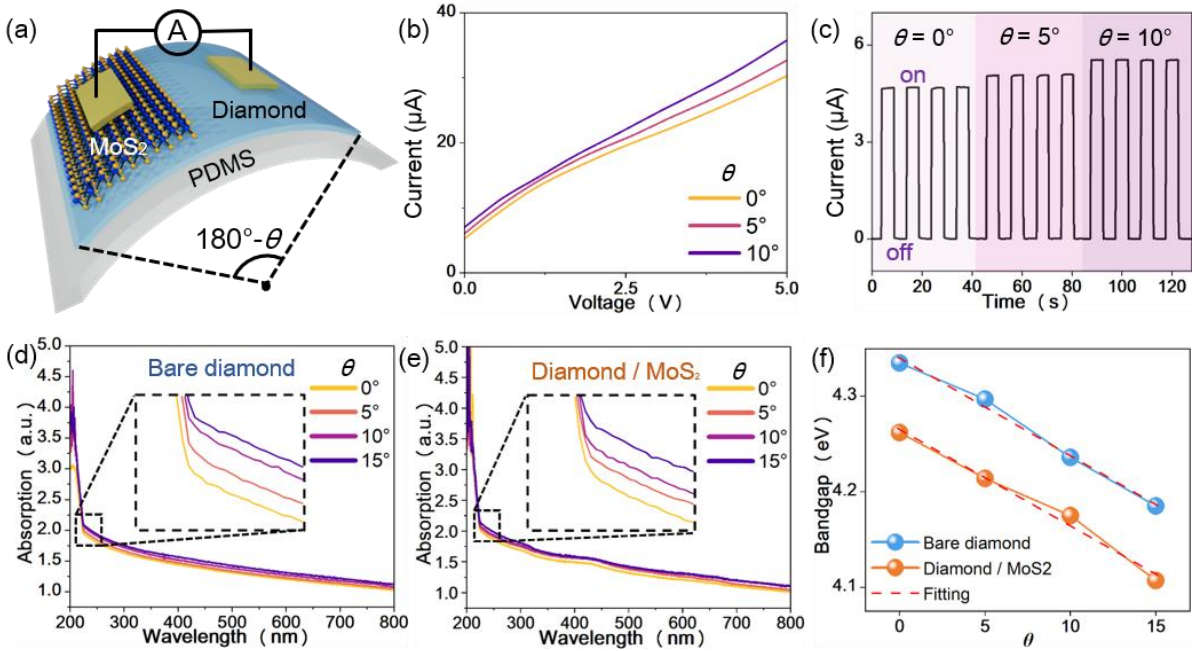

**Figure 3.** (a) Schematic illustration of the experimental setup for characterizing the flexible diamond UV detector under strain. (b) I-V curves of flexible diamond UV detector at varying values of $\theta$. (c) Photoresponse of the flexible diamond UV-detector during on/off loop of UV illumination at different values of $\theta$. Absorption spectra of (d) bare diamond and (e) diamond - $MoS_2$ integration measured at increasing $\theta$. (f) Bandgap of bare diamond and diamond integrated with $MoS_2$ as functions of bending angle $\theta$.

**Diamond UV imager**

To demonstrate the scalability of the proposed diamond UV detector, a proof-of-concept diamond UV imager is fabricated. Figure 4(a) presents the schematic illustration of the fabrication process of the diamond UV imager (details see section 5. Methods). (i) Diamond membrane with a thickness of 5 microns is mounted onto a silica substrate. (ii) $MoS_2$ monolayer (Sourced from SixCarbon Technology Shenzhen) is transferred onto diamond membrane. (iii)

Photolithography is applied to define stripes-shaped masks of MoS$_2$ for plasma etching, followed by photoresist removal. (iv) Second round of photolithography is applied to define the electrodes on the diamond – MoS$_2$ integration and silica substrate. (v) lift-off technique is used to remove the deposited gold. (vi) laser ablation isolates individual pixels on the ultrathin diamond membrane to minimize crosstalk. Figure 4(b) shows a photograph of the fabricated diamond UV imager. A magnified view of the unit cell, enclosed by the red dashed box in Figure 4(b), is exhibited in Figure 4(c), where the diamond – MoS$_2$ interface is highlighted with a white dashed line. Figure 4(d) presents a schematic drawing of the experimental setup used for characterizing the diamond UV imager, where the incident UV illumination from the UV light emitting diode (LED) is partially blocked by a hard mask. The custom letter masks ("H", "K", and "U") applied in the experiment and the corresponding photoresponse are shown in Figure 4 (e). The black and white area of the illustration drawing denotes areas with and without masks, respectively. The UV light exhibits peak intensity at the beam center, with a gradual decline in intensity from the center to the periphery, which coincides with the photoresponse of the UV imager. These results validate the detector's applicability for advanced applications in solar blind imaging systems and environmental monitoring, etc.

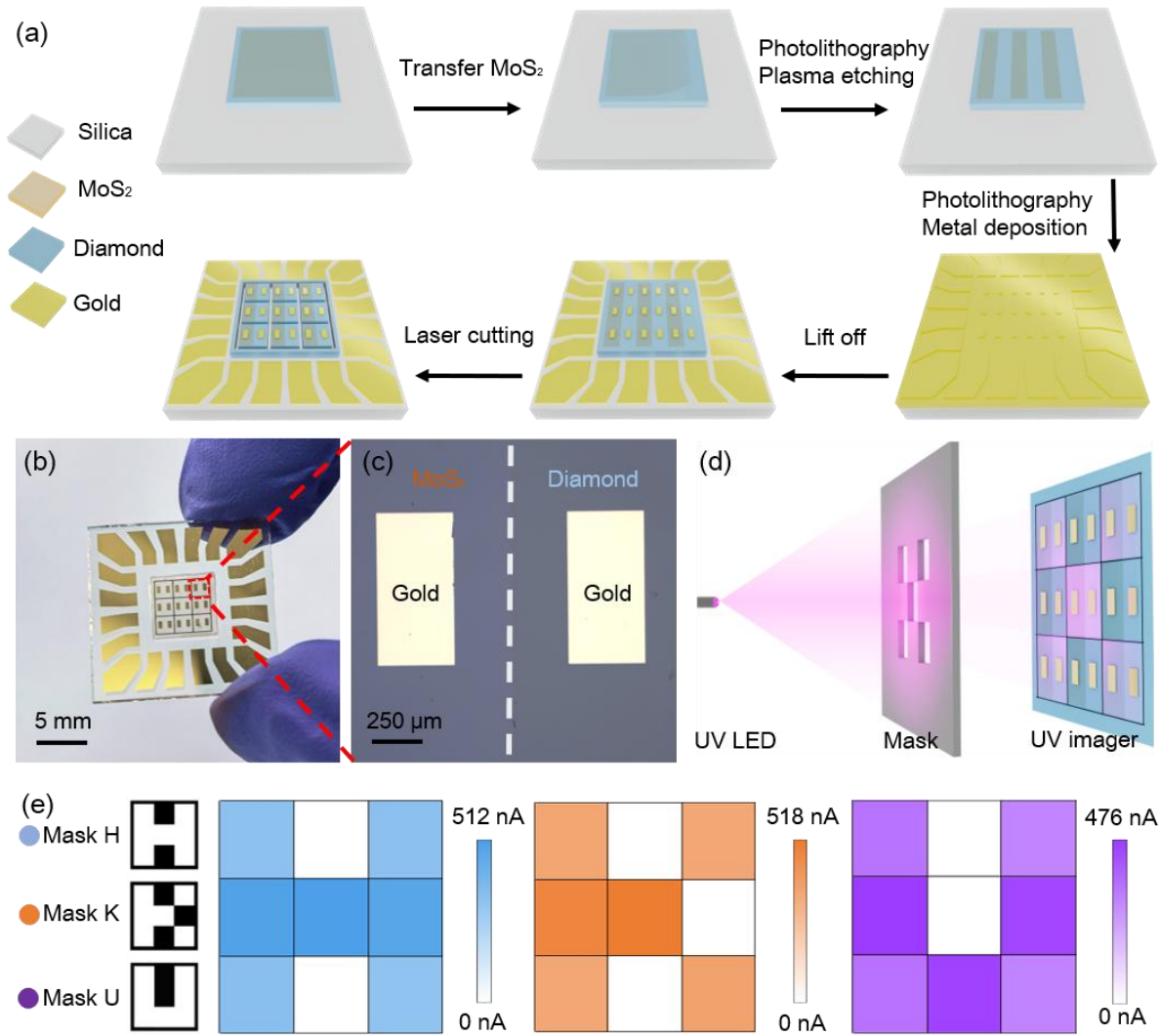

**Figure 4.** (a) Schematic illustration of the fabrication process of the diamond UV imager. (b) Photograph of the fabricated diamond UV imager. (c) Microscope image of a unit cell in the diamond UV imager. (d) Schematic illustration of experimental setup for measurement of diamond UV imager. (e) Illustration of masks and photoresponse of each pixel in diamond UV imager using letter masks "H", "K", and "U".

**Conclusion**

In this work, a large-scale, self-powered, flexible UV detector is demonstrated for the first time through heterogeneous integration of diamond and MoS$_2$. The device exhibits high sensitivity (94 mA·W$^{-1}$ at zero bias), high detectivity (5.88×10$^9$ Jones), and fast response time (40 ms). It also presents ultra-flexibility that is compatible with various curved and flexible platforms. Moreover, mechanical bending of the diamond membrane enables bandgap engineering, leading to a tunable photoresponse with an enhancement up to 117.9%, which are unattainable advantages over conventional diamond detectors. Furthermore, a diamond UV imager is carried out as a proof-of-concept demonstration, validating the detector's capability for spatially resolved UV detection. The combination of high optoelectronic performance, mechanical flexibility, and scalable fabrication positions this proposed UV detector as a promising candidate for next-generation UV sensing technologies, with potential applications in harsh environment monitoring, wearable devices, and advanced imaging systems.

**Methods**

*CVD growth of diamond membrane*

The heteroepitaxial growth of diamond membrane on a silicon (Si) substrate involves three essential steps: substrate pretreatment, diamond seed deposition, and chemical vapor deposition (CVD) of diamond membranes. Initially, substrate pretreatment began with hydrogen plasma treatment of the Si surface. A 2-inch silicon wafer was loaded into a microwave plasma-assisted chemical vapor deposition (MPCVD) reactor under 1300-W power, 35-Torr cavity pressure, and a 300-sccm hydrogen (H$_2$) gas flow for 10 minutes. Preparing the diamond seeds required a series of mixing, dispersing, and centrifuging steps. Diamond seeds (sourced from Tokyo Chemical Industry Company Limited) smaller than 10 nm were mixed with dimethyl sulfoxide (DMSO), anhydrous ethanol, and acetone at a mass ratio of 1:5000:250:250. This mixture was sonicated for 12 hours for dispersion, followed by centrifugation at 1000 rpm for 20 minutes to eliminate impurities. The suspension was then spin-coated onto the Si wafer under the following conditions: an initial spin at 500 rpm with three drops added within 15 seconds, followed by an increase to 4500 rpm for 110 seconds. This spin-coating process was repeated three times. Finally, the wafer with the deposited diamond seeds was placed into the MPCVD setup (Seki 6350) for diamond membrane growth. The primary parameters for growing a 5-μm-thick diamond membrane included 3400-W microwave power, a temperature of 900°C, a 15-sccm methane flow rate, and a 200-minute growth period.

*Preparation of flexible diamond UV detector*

Utilizing the edge-exposed exfoliation method, a 5-micron-thick diamond membrane with the size of 2 cm × 2 cm is exfoliated from the silicon substrate by sticky tape, which is then adhered onto a 1-mm-thick PDMS substrate. Monolayer $MoS_2$ on silicon substrate with 300 nm oxide layer is spin-coated with Polymethyl Methacrylate (PMMA) at the speed of 3000 rpm and baked at 150 °C for 120 seconds. Afterwards, 15 minutes of soaking in 20% potassium hydroxide (KOH) solution is used to dissolve the oxide layer. Therefore, the $MoS_2$ monolayer with PMMA is released in deionized water and transferred onto the diamond membrane. Subsequently, the assembly is oriented vertically and allowed to dry until no visible moisture remained. The assembly is then subjected to a 90°C annealing process for 30 minutes to promote covalent bonding at the diamond – $MoS_2$ interface. After removing the spin-coated PMMA using acetone, photoresist AZ-5214 is spin-coated at the speed of 3000 rpm and baked at 110 degrees Celsius for 90 seconds. Photolithography and lift-off techniques are applied to define the electrodes (5 nm of Chromium and 50 nm of gold) on diamond and MoS$_2$.

*Preparation of diamond UV imager*

After acquiring diamond – MoS$_2$ assembly spin-coated with photoresist following methods described in section 5.2. Photolithography is performed to define the stripes-shaped mask on MoS$_2$, which is therefore used as a mask during plasma etching. Using the same methods in section 5.2, electrodes (5 nm of Chromium and 50 nm of gold) on each unit cell of the UV imager and silica are fabricated. The diamond membrane is transferred onto the silica substrate using a thin layer of crystalbond 590-3. To avoid crosstalk between each unit cell, the laser is applied to cut the diamond membrane into separated unit cells.

**ACKNOWLEDGMENT**

Y. Wang, J. Jing, Y. Luo contributed equally to this work. K. H. Li acknowledges the Shenzhen Fundamental Research Program (JCYJ20220530113201003). Z. Chu acknowledges the financial support from the National Natural Science Foundation of China (NSFC) and the Research Grants Council (RGC) of the Hong Kong Joint Research Scheme (Project No. N_HKU750/23), HKU seed fund, and the Health@InnoHK program of the Innovation and Technology Commission of the Hong Kong SAR Government.

**Data Availability Statement**

The data that support the findings of this study are available from the corresponding author, upon reasonable request.

**Table of contents:**

By integrating diamond membrane with monolayer MoS2, large-scale flexible diamond UV detector with the thickness of 5 microns is achieved, which presents significantly higher responsivity than that of previous diamond UV detectors. By mechanically bending the device, tunable responsivity up to 117% is achieved. The cost-effectiveness, stability, self-power property, straightforward fabrication paves new way for next-generation UV detectors.

**Self-Powered, Ultra-thin, Flexible, and Scalable Ultraviolet Detector Utilizing Diamond-MoS$_2$ Heterojunction**

**TOC figure**

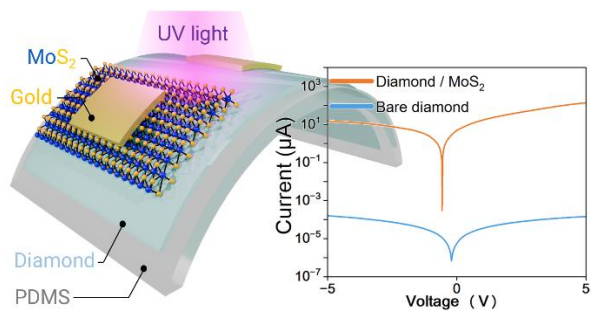